\documentclass[11pt]{article}
\usepackage{epsfig}
\usepackage{amsmath}
\usepackage{graphicx}
\usepackage{bbm}

\begin{document}

\author{S. Dev\thanks{dev5703@yahoo.com} $^{a, b}$, Radha Raman Gautam\thanks{gautamrrg@gmail.com} $^a$ and Lal Singh\thanks{lalsingh96@yahoo.com} $^a$}
\title{Hybrid Textures of the Right-Handed Majorana Neutrino Mass Matrix}
\date{$^a$\textit{Department of Physics, Himachal Pradesh University, Shimla 171005, INDIA.}\\
\smallskip
$^b$\textit{Department of Physics, School of Sciences, HNBG Central University, Srinagar, Uttarakhand 246174, INDIA.}}

\maketitle
\begin{abstract}
We perform a systematic study of neutrino mass matrices having a vanishing cofactor and an equality between two cofactors of the mass matrix. Such texture structures of the effective neutrino mass matrix arise from type-I seesaw mechanism when the Dirac neutrino mass matrix is diagonal with two equal elements and the right-handed Majorana neutrino mass matrix has hybrid textures with one equality of matrix elements and one zero matrix element. For three right-handed neutrinos there are sixty possible hybrid textures out of which only six are excluded by the present experimental data. We show that such textures can be derived using discrete symmetries. The predictions of experimentally allowed textures are examined for unknown parameters such as the effective Majorana mass of the electron neutrino and the Dirac-type CP-violating phase.  
\end {abstract}
%\pacs{14.60.Pq, 11.30.Hv, 14.60.St}

\section{Introduction}
During the past decade various neutrino oscillation experiments have determined the neutrino mass squared differences and the lepton mixing angles with a good precision \cite{experiments}. Especially the reactor mixing angle ($\theta_{13}$) has been measured rather precisely in recent experiments \cite{t2k, minos, dchooz, dayabay, reno}. The best fit value for $\theta_{13}$ is around $9^o$ which  has a large deviation from zero and provides an opportunity for the measurement of Dirac-type CP-violating phase ($\delta$) in the lepton mixing matrix. Several ideas have been mooted to accommodate the observed pattern of neutrino mass squared differences and mixing angles which include tribimaximal (TBM) mixing \cite{tbm} where one obtains fixed values of mixing angles and the neutrino masses are independent of the mixing angles. Such textures have been called mass independent textures \cite{low}. On the other hand,  zero textures \cite{zerotexture,xingzt2011} and vanishing minors \cite{zerominor, ourzerominor} which relate neutrino masses with mixing angles are called mass-dependent textures. Hybrid textures which imply one equality between matrix elements and one zero element in the effective neutrino mass matrix also fall in the category of mass dependent textures and have been studied earlier in a basis where the charged lepton mass matrix is diagonal \cite{tanimoto, hybrid, zhouhybrid} and in a basis where both the charged lepton and the neutrino mass matrix have hybrid textures \cite{hybridnondiagonal}. In the present work, we examine the phenomenological implications of hybrid textures of the inverse neutrino mass matrix in a basis where the charged lepton mass matrix is diagonal. \\ 
To understand the smallness of neutrino masses as compared to charged fermion masses, the seesaw mechanism \cite{seesaw1} is regarded as the prime candidate. In the framework of type-I seesaw mechanism, the effective Majorana neutrino mass matrix is given by 
\begin{equation}
M_\nu \approx -M_D M_R^{-1} M_D^T
\end{equation}
where $M_D$ is the Dirac neutrino mass matrix and $M_R$ is the right-handed Majorana neutrino mass matrix. Within this framework, $M_\nu$ is a quantity derived from $M_D$ and $M_R$. Thus, in the context of type-I seesaw mechanism, the zeros and equalities of $M_D$ and $M_R$ are more fundamental. In a basis where $M_D$ is diagonal, the zero textures of $M_R$ show as vanishing cofactors in $M_\nu$ \cite{zerominor} and the equalities in $M_R$ propagate as equal cofactors in $M_\nu$ provided that there are equal non-zero elements in diagonal $M_D$ \cite{ourtec}. We consider another possibility where $M_R$ has hybrid textures in a basis where $M_D$ is diagonal with one equality between diagonal elements. Such $M_D$ and $M_R$ give rise to the neutrino mass matrix having one equality between the cofactors and one vanishing cofactor. The equality between the cofactors and the vanishing cofactor in $M_\nu$ correspond to the equal and zero elements of $M_R$, respectively. Such textures can also be seen as hybrid textures of the inverse $M_\nu$. Thus, effectively we are studying the hybrid textures of $M_\nu^{-1}$. For three right-handed neutrinos, there are sixty possible hybrid textures of $M_R$ which are listed in Table 1. To illustrate how such textures can be realised we present a flavor model for class $IA$ using discrete flavor symmetries. 
\begin{table}[!]
\begin{small}
\noindent\makebox[\textwidth]{
\begin{tabular}{|c|c|c|c|c|c|c|}
\hline  & $A$ & $B$ & $C$ & $D$ & $E$ & $F$  \\
\hline $I$ & $\left(
\begin{array}{ccc}
0 & b & b \\  b & d & e \\ b & e & f
\end{array}
\right)$ & $\left(
\begin{array}{ccc}
a & 0 & a \\  0 & d & e \\ a & e & f
\end{array}
\right)$ & $\left(
\begin{array}{ccc}
a & a & 0 \\  a & d & e \\ 0 & e & f
\end{array}
\right)$ & $\left(
\begin{array}{ccc}
a & a & c \\  a & 0 & e \\ c & e & f
\end{array}
\right)$&$\left(
\begin{array}{ccc}
a & a & c \\  a & d & 0 \\ c & 0 & f
\end{array}
\right)$& $\left(
\begin{array}{ccc}
a & a & c \\  a & d & e \\ c & e & 0
\end{array}
\right)$\\

\hline $II$ & $\left(
\begin{array}{ccc}
0 & b & c \\  b & b & e \\ c & e & f
\end{array}
\right)$ & $\left(
\begin{array}{ccc}
a & 0 & c \\  0 & a & e \\ c & e & f
\end{array}
\right)$ & $\left(
\begin{array}{ccc}
a & b & 0 \\  b & a & e \\ 0 & e & f
\end{array}
\right)$  & $\left(
\begin{array}{ccc}
a & b & a \\  b & 0 & e \\ a & e & f
\end{array}
\right)$ & $\left(
\begin{array}{ccc}
a & b & a \\  b & d & 0 \\ a & 0 & f
\end{array}
\right)$ & $\left(
\begin{array}{ccc}
a & b & a \\  b & d & e \\ a & e & 0
\end{array}
\right)$ \\

\hline $III$ & $\left(
\begin{array}{ccc}
0 & b & c \\  b & d & b \\ c & b & f
\end{array}
\right)$ & $\left(
\begin{array}{ccc}
a & 0 & c \\  0 & d & a \\ c & a & f
\end{array}
\right)$  & $\left(
\begin{array}{ccc}
a & b & 0 \\  b & d & a \\ 0 & a & f
\end{array}
\right)$ & $\left(
\begin{array}{ccc}
a & b & c \\  b & 0 & a \\ c & a & f
\end{array}
\right)$  & $\left(
\begin{array}{ccc}
a & b & c \\  b & a & 0 \\ c & 0 & f
\end{array}
\right)$ & $\left(
\begin{array}{ccc}
a & b & c \\  b & a & e \\ c & e & 0
\end{array}
\right)$ \\

\hline $IV$ & $\left(
\begin{array}{ccc}
0 & b & c \\  b & d & e \\ c & e & b
\end{array}
\right)$ & $\left(
\begin{array}{ccc}
a & 0 & c \\  0 & b & e \\ c & e & a
\end{array}
\right)$ &$\left(
\begin{array}{ccc}
a & b & 0 \\  b & d & e \\ 0 & e & a
\end{array}
\right)$& $\left(
\begin{array}{ccc}
a & b & c \\  b & 0 & e \\ c & e & a
\end{array}
\right)$ &$\left(
\begin{array}{ccc}
a & b & c \\  b & d & 0 \\ c & 0 & a
\end{array}
\right)$ & $\left(
\begin{array}{ccc}
a & b & c \\  b & d & a \\ c & a & 0
\end{array}
\right)$\\

\hline $V$ & $\left(
\begin{array}{ccc}
0 & b & c \\  b & c & e \\ c & e & f
\end{array}
\right)$  & $\left(
\begin{array}{ccc}
a & 0 & c \\  0 & c & e \\ c & e & f
\end{array}
\right)$ &$\left(
\begin{array}{ccc}
a & b & 0 \\  b & b & e \\ 0 & e & f
\end{array}
\right)$  & $\left(
\begin{array}{ccc}
a & b & b \\  b & 0 & e \\ b & e & f
\end{array}
\right)$ & $\left(
\begin{array}{ccc}
a & b & b \\  b & d & 0 \\ b & 0 & f
\end{array}
\right)$ & $\left(
\begin{array}{ccc}
a & b & b \\  b & d & e \\ b & e & 0
\end{array}
\right)$ \\

\hline $VI$ & $\left(
\begin{array}{ccc}
0 & b & c \\ b & d & c \\ c & c & f
\end{array}
\right)$ & $\left(
\begin{array}{ccc}
a & 0 & c \\  0 & d & c \\ c & c & f
\end{array}
\right)$ &$\left(
\begin{array}{ccc}
a & b & 0 \\  b & d & b \\ 0 & b & f
\end{array}
\right)$  &$\left(
\begin{array}{ccc}
a & b & c \\  b & 0 & b \\ c & b & f
\end{array}
\right)$ & $\left(
\begin{array}{ccc}
a & b & c \\  b & b & 0 \\ c & 0 & f
\end{array}
\right)$ &$\left(
\begin{array}{ccc}
a & b & c \\  b & b & e \\ c & e & 0
\end{array}
\right)$  \\

\hline $VII$ & $\left(
\begin{array}{ccc}
0 & b & c \\ b & d & e \\ c & e & c
\end{array}
\right)$ & $\left(
\begin{array}{ccc}
a & 0 & c \\  0 & d & e \\ c & e & c
\end{array}
\right)$ & $\left(
\begin{array}{ccc}
a & b & 0 \\  b & d & e \\ 0 & e & b
\end{array}
\right)$&$\left(
\begin{array}{ccc}
a & b & c \\  b & 0 & e \\ c & e & b
\end{array}
\right)$ & $\left(
\begin{array}{ccc}
a & b & c \\  b & d & 0 \\ c & 0 & b
\end{array}
\right)$ & $\left(
\begin{array}{ccc}
a & b & c \\  b & d & b \\ c & b & 0
\end{array}
\right)$  \\

\hline $VIII$ & $\left(
\begin{array}{ccc}
0 & b & c \\ b & d & d \\ c & d & f
\end{array}
\right)$ & $\left(
\begin{array}{ccc}
a & 0 & c \\  0 & d & d \\ c & d & f
\end{array}
\right)$ &$\left(
\begin{array}{ccc}
a & b & 0 \\  b & d & d \\ 0 & d & f
\end{array}
\right)$  & $\left(
\begin{array}{ccc}
a & b & c \\  b & 0 & c \\ c & c & f
\end{array}
\right)$ & $\left(
\begin{array}{ccc}
a & b & c \\  b & c & 0 \\ c & 0 & f
\end{array}
\right)$ & $\left(
\begin{array}{ccc}
a & b & c \\  b & c & e \\ c & e & 0
\end{array}
\right)$  \\

\hline $IX$ & $\left(
\begin{array}{ccc}
0 & b & c \\ b & d & e \\ c & e & d
\end{array}
\right)$ & $\left(
\begin{array}{ccc}
a & 0 & c \\  0 & d & e \\ c & e & d
\end{array}
\right)$&$\left(
\begin{array}{ccc}
a & b & 0 \\  b & d & e \\ 0 & e & d
\end{array}
\right)$ & $\left(
\begin{array}{ccc}
a & b & c \\  b & 0 & e \\ c & e & c
\end{array}
\right)$ & $\left(
\begin{array}{ccc}
a & b & c \\  b & d & 0 \\ c & 0 & c
\end{array}
\right)$ & $\left(
\begin{array}{ccc}
a & b & c \\  b & d & c \\ c & c & 0
\end{array}
\right)$  \\

\hline $X$ & $\left(
\begin{array}{ccc}
0 & b & c \\ b & d & e \\ c & e & e
\end{array}
\right)$ &$\left(
\begin{array}{ccc}
a & 0 & c \\  0 & d & e \\ c & e & e
\end{array}
\right)$&$\left(
\begin{array}{ccc}
a & b & 0 \\  b & d & e \\ 0 & e & e
\end{array}
\right)$ &$\left(
\begin{array}{ccc}
a & b & c \\  b & 0 & e \\ c & e & e
\end{array}
\right)$ & $\left(
\begin{array}{ccc}
a & b & c \\  b & d & 0 \\ c & 0 & d
\end{array}
\right)$ & $\left(
\begin{array}{ccc}
a & b & c \\  b & d & d \\ c & d & 0
\end{array}
\right)$  \\

\hline
\end{tabular}}
\end{small}
\caption{Sixty possible hybrid texture structures of $M_{R}$.}
\end{table}

\section{The Neutrino Mass Matrix}
Assuming neutrinos to be Majorana particles, we reconstruct the neutrino mass matrix in the flavor basis (where the charged lepton mass matrix $``M_l"$ is diagonal). In this basis, the complex symmetric neutrino mass matrix is diagonalized by a unitary matrix $V'$ as
\begin{equation}
M_{\nu}= V' M_{\nu}^{diag}V'^{T}
\end{equation}
where $M_{\nu}^{diag}$ = diag$(m_1,m_2,m_3)$. \\
The unitary matrix $V'$ can be parametrized as
\begin{equation}
V' = P_lV \ \ \ \textrm{with}\ \ \ \ V = UP_\nu
\end{equation}
where  \cite{foglipdg}
\begin{equation}
U= \left(
\begin{array}{ccc}
c_{12}c_{13} & s_{12}c_{13} & s_{13}e^{-i\delta} \\
-s_{12}c_{23}-c_{12}s_{23}s_{13}e^{i\delta} &
c_{12}c_{23}-s_{12}s_{23}s_{13}e^{i\delta} & s_{23}c_{13} \\
s_{12}s_{23}-c_{12}c_{23}s_{13}e^{i\delta} &
-c_{12}s_{23}-s_{12}c_{23}s_{13}e^{i\delta} & c_{23}c_{13}
\end{array}
\right)
\end{equation} with $s_{ij}=\sin\theta_{ij}$ and $c_{ij}=\cos\theta_{ij}$ and
\begin{small}
\begin{center}
$P_\nu = \left(
\begin{array}{ccc}
1 & 0 & 0 \\ 0 & e^{i\alpha} & 0 \\ 0 & 0 & e^{i(\beta+\delta)}
\end{array}
\right)$ , \ \ \ 
$P_l = \left(
\begin{array}{ccc}
e^{i\varphi_e} & 0 & 0 \\ 0 & e^{i\varphi_\mu} & 0 \\ 0 & 0 & e^{i\varphi_\tau}
\end{array}
\right).$
\end{center}
\end{small}
$P_\nu$ is the diagonal phase matrix with 
two Majorana-type CP-violating phases $\alpha$, $\beta$ and one Dirac-type CP-violating phase $\delta$. The phase matrix $P_l$ is not observable and depends on the phase convention. The matrix $V$ is called the neutrino mixing matrix or the Pontecorvo-Maki-Nakagawa-Sakata (PMNS) matrix  \cite{pmns}. Using Eq. (2) and Eq. (3), the neutrino mass matrix can be written as
\begin{equation}
M_{\nu}=P_l U P_\nu M_{\nu}^{diag}P_\nu^{T}U^{T}P_l^T.
\end{equation}
The CP-violation in neutrino oscillation experiments can be described through a rephasing invariant quantity, $J_{CP}$ \cite{jarlskog} with $J_{CP}=Im(U_{e1}U_{\mu2}U_{e2}^*U_{\mu1}^*)$. In the above parametrization, $J_{CP}$ is given by
\begin{equation}
J_{CP} = s_{12}s_{23}s_{13}c_{12}c_{23}c_{13}^2 \sin \delta \   .
\end{equation}

\subsection{Neutrino Mass Matrices with an Equality Between the Cofactors and a Vanishing Cofactor}
In the effective neutrino mass matrix, the simultaneous existence of an equality between two cofactors and one vanishing cofactor implies
\begin{align}
&(-1^{(\gamma \xi)})(  e^{i(\varphi_a +\varphi_b + \varphi_c +\varphi_d)}M_{\nu (ab)} M_{\nu (cd)}- e^{i(\varphi_f +\varphi_g + \varphi_m +\varphi_n)}M_{\nu
(fg)} M_{\nu (mn)})-  \nonumber \\  &(-1^{(\zeta \eta)})( e^{i(\varphi_p +\varphi_q + \varphi_r +\varphi_s)}M_{\nu (pq)} M_{\nu (rs)}- e^{i(\varphi_t +\varphi_u + \varphi_v +\varphi_w)}M_{\nu
(tu)} M_{\nu (vw)})=0 \\
& e^{i(\varphi_{a'} +\varphi_{b'} + \varphi_{c'} +\varphi_{d'})}M_{\nu (a'b')} M_{\nu (c'd')}- e^{i(\varphi_{f'} +\varphi_{g'} + \varphi_{m'} +\varphi_{n'})}M_{\nu
(f'g')} M_{\nu (m'n')}=0 .
\end{align}
The condition for equality between two cofactors in the neutrino mass matrix [Eq. (7)] can be written as
\begin{align}
&(-1^{(\gamma \xi)})( Q_1 M_{\nu (ab)} M_{\nu (cd)}- Q_2 M_{\nu
(fg)} M_{\nu (mn)})- \nonumber \\ &(-1^{(\zeta \eta)})( Q_3 M_{\nu (pq)} M_{\nu (rs)}- Q_4 M_{\nu
(tu)} M_{\nu (vw)})=0 . 
\end{align}
It is inherent in the properties of cofactors that when we substitute $\varphi_j, (j = e, \mu, \tau$), then $Q_1 = Q_2$ and $Q_3 = Q_4$. Thus, we have
\begin{align}
&(-1^{(\gamma \xi)})Q_1( M_{\nu (ab)} M_{\nu (cd)}- M_{\nu
(fg)} M_{\nu (mn)})- \nonumber \\ & (-1^{(\zeta \eta)})Q_3( M_{\nu (pq)} M_{\nu (rs)}- M_{\nu
(tu)} M_{\nu (vw)})=0 .  
\end{align}
or
\begin{align}
&(-1^{(\gamma \xi)})Q( M_{\nu (ab)} M_{\nu (cd)}- M_{\nu
(fg)} M_{\nu (mn)})- \nonumber \\ & (-1^{(\zeta \eta)})( M_{\nu (pq)} M_{\nu (rs)}- M_{\nu
(tu)} M_{\nu (vw)})=0 . 
\end{align}
where $Q = \frac{Q_1}{Q_3}$.\\
The two conditions Eq. (8) and Eq. (11) take the following form when expressed in terms of the mixing matrix elements and mass eigenvalues:
\begin{align}
\sum_{k,l=1}^{3} & \{(-1^{(\gamma \xi)})Q(V_{ak}V_{bk}V_{cl}V_{dl}-V_{fk}V_{gk}V_{ml}V_{nl})- \nonumber \\ &(-1^{(\zeta \eta)})(V_{pk}V_{qk}V_{rl}V_{sl}-V_{tk}V_{uk}V_{vl}V_{wl})\}m_{k}m_{l}=0 , \\ 
\sum_{k,l=1}^{3}&(V_{a'k}V_{b'k}V_{c'l}V_{d'l}-V_{f'k}V_{g'k}V_{m'l}V_{n'l})m_{k}m_{l}=0 \ .
\end{align}
The above equations can be rewritten as
\begin{align}
m_1 m_2 A_3e^{2i\alpha} + m_2 m_3 A_1e^{2i(\alpha+\beta +\delta )}+m_3 m_1A_2e^{2i(\beta +\delta)}&=0 \ , \\
m_1 m_2 B_3e^{2i\alpha} + m_2 m_3 B_1e^{2i(\alpha+\beta +\delta )}+m_3 m_1 B_2e^{2i(\beta +\delta)}&=0 \ 
\end{align}
where
\begin{align}
A_h & =(-1^{(\gamma \xi)})Q(U_{ak}U_{bk}U_{cl}U_{dl}-U_{fk}U_{gk}U_{ml}U_{nl})- \nonumber \\ & (-1^{(\zeta \eta)})(U_{pk}U_{qk}U_{rl}U_{sl}-U_{tk}U_{uk}U_{vl}U_{wl})+(k\leftrightarrow l) , \\
B_h & =(U_{a'k}U_{b'k}U_{c'l}U_{d'l}-U_{f'k}U_{g'k}U_{m'l}U_{n'l})+(k\leftrightarrow l) 
\end{align}
with $(h,k,l)$ as the cyclic permutation of (1,2,3). The two complex Eqs. (14) and (15) contain nine physical parameters which are the three neutrino masses ($m_{1}$, $m_{2}$, $m_{3}$), the three mixing angles ($\theta _{12}$, $\theta _{23}$, $\theta _{13}$) and three CP-violating phases ($\alpha $, $\beta $, $\delta $). In addition, there are three unobservable phases $(\varphi_e, \varphi_\mu, \varphi_\tau)$ which enter in the mass ratios as a phase difference. The masses $m_{2}$ and $m_{3}$ can be calculated from the mass-squared differences $\Delta m_{21}^{2}$ ($\Delta m_{21}^{2} \equiv m_2^2 - m_1^2$) and $|\Delta m_{32}^{2}|$ using the following relations
\begin{equation}
m_{2}=\sqrt{m_{1}^{2}+\Delta m_{21}^{2}} \ , \ \  m_{3}=\sqrt{m_{2}^{2}+|\Delta m_{32}^{2}|} \ 
\end{equation}
with $m_2 > m_3$ for an Inverted Spectrum (IS) and $m_2 < m_3$ for the Normal Spectrum (NS). 
By using the experimental inputs of the two mass-squared differences and the three mixing angles, we can constrain the other parameters. Simultaneously solving Eqs. (14) and (15) for the two mass ratios, we obtain

\begin{small}
\begin{align}
\frac{m_1}{m_2}e^{-2i\alpha }&=\frac{A_3 B_2 - A_2 B_3}{A_1 B_3 - A_3 B_1} ,\\
\frac{m_1}{m_3}e^{-2i\beta }&=\frac{A_2 B_3 - A_3 B_2 }{A_1 B_2- A_2 B_1}e^{2i\delta } \ .
\end{align}
\end{small}
We denote the magnitudes of the above two mass ratios by
\begin{align}
\rho &=\left|\frac{m_1}{m_3}e^{-2i\beta }\right| ,\\
\sigma &=\left|\frac{m_1}{m_2}e^{-2i\alpha }\right|.
\end{align}
The Majorana-type CP-violating phases $\alpha$ and $\beta$ are given by
 \begin{small}
\begin{align}
\alpha & =-\frac{1}{2}\textrm{arg}\left(\frac{A_3 B_2 - A_2 B_3}{A_1 B_3 - A_3 B_1}\right), \\
\beta & =-\frac{1}{2}\textrm{arg}\left(\frac{A_2 B_3 - A_3 B_2 }{A_1 B_2- A_2 B_1}e^{2i\delta }\right).
\end{align}
\end{small}
Since, $\Delta m_{21}^{2}$ and $|\Delta m_{32}^{2}|$ are experimentally known, the two mass ratios $(\rho,\sigma)$ in Eqs. (21) and (22) can be used to calculate $m_1$.
This can be done by inverting Eq. (18) to obtain the two values of $m_1$, viz.
\begin{small}
\begin{equation}
m_{1}=\sigma \sqrt{\frac{ \Delta
m_{21}^{2}}{1-\sigma ^{2}}} \ , \ \ 
m_{1}=\rho \sqrt{\frac{\Delta m_{21}^{2}+
|\Delta m_{32}^{2}|}{ 1-\rho^{2}}} .
\end{equation}
\end{small}
There exists a permutation symmetry between different patterns of two-zero textures \cite{xingzt2011}, in the case of two vanishing minors \cite{ourzerominor} and in the case of two equalities in $M_\nu$ \cite{ourtec}. Similarly, there exists a permutation symmetry between different hybrid textures of $M_\nu^{-1}$. The permutation matrix is given by
\begin{small} 
\begin{equation}
P_{23} = \left(
\begin{array}{ccc}
1&0&0\\
0&0&1\\
0&1&0\\
\end{array}
\right).
\end{equation}
\end{small}
For example, the neutrino mass matrix for class $IC$ can be obtained from class $IB$ by the transformation
\begin{equation}
M_{\nu}^{IC} = P_{23}M_{\nu}^{IB}P_{23}^T \ .
\end{equation}
This leads to the following relations between the parameters of the classes related by the $2$-$3$ permutation symmetry: 
\begin{equation}
\theta_{12}^{IC} = \theta_{12}^{IB}, \ \theta_{13}^{IC} = \theta_{13}^{IB}, \ \theta_{23}^{IC} = \frac{\pi}{2}-\theta_{23}^{IB}, \ \delta^{IC} = \delta^{IB} - \pi \ .
\end{equation}
The textures related by the 2-3 permutation symmetry are:
\begin{small}
\begin{align}
& IB \leftrightarrow IC, \ ID \leftrightarrow IIF, \ IE \leftrightarrow IIE, \ IF \leftrightarrow IID, \ IIA \leftrightarrow VIIA, \ IIB \leftrightarrow IVC, \ IIC \leftrightarrow IVB, \nonumber \\  & IIIA \leftrightarrow VIA, \ IIIB \leftrightarrow IIIC, \ IIID \leftrightarrow IVF, \ IIIE \leftrightarrow IVE, \ IIIF \leftrightarrow IVD, \ IVA \leftrightarrow VA, \nonumber  \\ 
& VB \leftrightarrow VIIC, \ VC \leftrightarrow VIIB, \  VD \leftrightarrow VF, \ VIB \leftrightarrow VIC, \ VID \leftrightarrow IXF, \ VIE \leftrightarrow IXE, \nonumber  \\ &  VIF \leftrightarrow IXD, \ VIID \leftrightarrow VIIIF, \ VIIE \leftrightarrow VIIIE, \ VIIF \leftrightarrow VIIID, \ VIIIA \leftrightarrow XA, \nonumber  \\ 
&  VIIIB \leftrightarrow XC, \ VIIIC \leftrightarrow XB, \ IXB \leftrightarrow IXC, \ XD \leftrightarrow XF. 
\end{align}
\end{small}
The remaining textures namely:
\begin{small}
\begin{align}
IA, \ VE, \ IXA, \ XE
\end{align}
\end{small}
transform unto themselves.

\section{Numerical Analysis and Results}
The latest global fit results on neutrino oscillation parameters at 1$\sigma$, 2$\sigma$ and 3$\sigma$ \cite{valledata} are summarised in Table 2.
The effective Majorana mass of the electron neutrino ($M_{ee}$) which determines the rate of neutrinoless double beta (NDB) decay is given by
\begin{equation}
M_{ee}= |m_1c_{12}^2c_{13}^2+ m_2s_{12}^2c_{13}^2 e^{2i\alpha}+ m_3s_{13}^2e^{2i\beta}|.
\end{equation}
Observation of NDB decay will imply that neutrinos are Majorana fermions. NDB decay also provides a way to probe the neutrino mass scale. There are a large number of projects such as CUORICINO \cite{cuoricino}, CUORE \cite{cuore}, GERDA \cite{gerda}, MAJORANA \cite{majorana}, SuperNEMO \cite{supernemo}, EXO \cite{exo}, GENIUS \cite{genius} which aim to achieve a sensitivity upto 0.01 eV for $M_{ee}$. We take the upper limit of $M_{ee}$ to be 0.5 eV \cite{rodejohann}. In addition, cosmological observations put an upper bound on the sum of light neutrino masses
\begin{equation}
\Sigma = \sum_{i=1}^3 m_i \ .
\end{equation} 
Recent data from Planck satellite \cite{planck} combined with other cosmological data limit $\Sigma < 0.23$ at $95\%$ Confidence level. However, these bounds are strongly dependent on model details and the data set used. Thus, in our numerical analysis we take the upper limit on  $\Sigma$ to be 1 eV.\\
In our numerical programs, the constraints implied by an equality between the two cofactors of $M_\nu$ and one vanishing cofactor in $M_\nu$ are used by equating the two values of $m_1$ obtained in Eq. (25). These two values of $m_1$ should be equal to within the errors of the oscillation parameters. We vary the known oscillation parameters randomly within their 3$\sigma$ experimental ranges given in \cite{valledata}. The unknown Dirac-type CP-violating phase $\delta$ is varied randomly within its full possible range. For the numerical analysis, we generate $10^7$ random points ( $10^8$ when the number of allowed points is small) for the allowed 3$\sigma$ ranges of oscillation parameters.\\
\begin{table}[!]
\begin{center}
\begin{tabular}{|c|c|}
\hline Parameter & Mean $^{(+1 \sigma, +2 \sigma, +3 \sigma)}_{(-1 \sigma, -2 \sigma, -3 \sigma)}$ \\
\hline $\Delta m_{21}^{2} [10^{-5}eV^{2}]$ & $7.62_{(-0.19,-0.35,-0.5)}^{(+0.19,+0.39,+0.58)}$ \\ 
\hline $\Delta m_{31}^{2} [10^{-3}eV^{2}]$ & $2.55_{(-0.09,-0.19,-0.24)}^{(+0.06,+0.13,+0.19)}$, \\&
$(-2.43_{(-0.07,-0.15,-0.21)}^{(+0.09,+0.19,+0.24)})$ \\ 
\hline $\sin^2 \theta_{12}$ & $0.32_{(-0.017,-0.03,-0.05)}^{(+0.016,+0.03,+0.05)}$ \\ 
\hline $\sin^2 \theta_{23}$ & $0.613_{(-0.04,-0.233,-0.25)}^{(+0.022,+0.047,+0.067)}$, \\& $(0.60_{(-0.031,-0.210,-0.230)}^{(+0.026,+0.05,+0.07)})$ \\ 
\hline $\sin^2 \theta_{13}$ & $0.0246_{(-0.0029,-0.0054,-0.0084)}^{(+0.0028,+0.0056,+0.0076)}$,\\& $(0.0250_{(-0.0027,-0.005,-0.008)}^{(+0.0026,+0.005,+0.008)})$ \\ 
\hline 
\end{tabular}
\caption{Current Neutrino oscillation parameters from global fits \cite{valledata}. The upper (lower) row corresponds to Normal (Inverted) Spectrum, with $\Delta m^2_{31} > 0$ ($\Delta m^2_{31} < 0$).}
\end{center}
\end{table}
%\newpage
\begin{table}[!h]
\centering
\resizebox{5. in}{4.2 in}{\begin{tabular}{|c|c|c|c|c|c|}
 \hline 
Texture    & Spectrum   & $M_{ee}$ (eV)           & $m_{o}(eV)$ & $\theta_{23}$ & Majorana Phases         \\
 \hline
IA         & IS         & $0.01$-$0.055$          &  $0.0007$   & -    &  $\beta$ = $70^{\circ}$-$110^{\circ}$\\
 \hline 
IB (IC)    & IS         & $0.034$-$0.05$          &  $0.007$    & -    &  $\alpha$=$25^{\circ}$-$55^{\circ}$,$125^{\circ}$-$155^{\circ}$   \\
 \hline
ID (IIF)   & NS         & $0$-$0.15$              &  $0.003$    &-    &  $\alpha$=$50^{\circ}$-$130^{\circ}$              \\ 
           & IS         & $0.03$-$0.20$           &  $0.03$    &$<45^{\circ}$($> 45^{\circ}$)     &  $\alpha$=$45^{\circ}$-$165^{\circ}$,$115^{\circ}$-$135^{\circ}$  \\
\hline  
IE (IIE)   & NS         & $0$-$0.07$              &  $0.002$    &-                                 &  $\alpha$=$40^{\circ}$-$140^{\circ}$ \\
\hline
IF (IID)   & NS         & $0$-$0.16$           &  $0.002$   &-      &  $\alpha$=$50^{\circ}$-$130^{\circ}$  \\
           & IS         & $0.02$-$0.16$     &  $0.06$   &$>45^{\circ}$($< 45^{\circ}$)       &  $\alpha$=$60^{\circ}$-$120^{\circ}$  \\
 \hline
IIB (IVC)  & NS         & $0.04$-$0.35$           &  $0.05$   &-      &  $\alpha$=$0^{\circ}$-$30^{\circ}$,~$150^{\circ}$-$180^{\circ}$ \\ 
           & IS         & $0.04$-$0.35$           &  $0.008$   &-      &  $\alpha,~\beta$=$0^{\circ}$-$30^{\circ}$,~$150^{\circ}$-$180^{\circ}$ \\
 \hline
IIC (IVB)  & IS         & $0.02$-$0.35$           &  $0.01$   &-     &  $\alpha$=$0^{\circ}$-$80^{\circ}$,~$100^{\circ}$-$180^{\circ}$ \\
\hline 
IIIB (IIIC) & IS        & $0.03$-$0.30$           & $0.015$    &-      &  $\alpha$=$0^{\circ}$-$60^{\circ}$,~$120^{\circ}$-$180^{\circ}$ \\
\hline 
IIID (IVF) & NS         & $0.002$-$0.25$          & $0.015$    &-      &  -                              \\
\hline
IIIE (IVE) & NS         & $0$-$0.35$              & $0.01$   &-      &  $\alpha$=$5^{\circ}$-$175^{\circ}$  \\
\hline
IIIF (IVD) & NS         & $0$-$0.16$              & $0.01$     &-      &  $\alpha$=$60^{\circ}$-$120^{\circ}$   \\
           & IS   & $0.04$-$0.18$            & $0.02$     &$>45^{\circ}$($< 45^{\circ}$) &  $\alpha$=$0^{\circ}$-$80^{\circ}$,~$100^{\circ}$-$180^{\circ}$ \\
\hline  
VB (VIIC)  & NS        & $0.004$-$0.35$          & $0.005$   &$>45^{\circ}$($< 45^{\circ}$)      &  $\alpha$=$0^{\circ}$-$20^{\circ}$,~$160^{\circ}$-$180^{\circ}$  \\ 
           & IS         & $0.04$-$0.35$           & $0.010$    &$<45^{\circ}$($> 45^{\circ}$)      &  $\alpha$=$0^{\circ}$-$20^{\circ}$,~$160^{\circ}$-$180^{\circ}$ \\
 \hline
VC (VIIB)   & NS        & $0.005$-$0.30$          & $0.007$   &$>45^{\circ}$($< 45^{\circ}$)      &  $\alpha$=$0^{\circ}$-$10^{\circ}$,~$170^{\circ}$-$180^{\circ}$ \\
            & IS        & $0.01$-$0.30$              & $0.01$   &$<45^{\circ}$($> 45^{\circ}$)       &   -        \\
\hline
VD (VF)    & NS         & $0$-$0.35$              & $0.001$    &-       &  -    \\
           & IS         & $0.04$-$0.35$           & $0.020$   &-       &  $\alpha$=$0^{\circ}$-$45^{\circ}$,~$135^{\circ}$-$180^{\circ}$ \\
 \hline
VE         & NS         & $0$-$0.20$              & $0.001$   &-       &  -                                \\
           & IS         & $0.1$-$0.35$            & $0.10$    &-      &  $\alpha$=$0^{\circ}$-$10^{\circ}$,
~$170^{\circ}$-$180^{\circ}$,~$\beta$=$20^{\circ}$-$160^{\circ}$ \\
\hline
VIB (VIC)   & NS         & $0.002$-$0.35$         & $0.004$ &-        &  $\alpha$=$0^{\circ}$-$20^{\circ}$,~$160^{\circ}$-$180^{\circ}$\\
            & IS         & $0.02$-$0.35$          & $0.013$  &-        &  $\alpha$=$0^{\circ}$-$80^{\circ}$,~$100^{\circ}$-$180^{\circ}$ \\
\hline
VID (IXF)  & NS         & $0.005$-$0.12$          & $0.006$  &-        &  $\alpha$=$0^{\circ}$-$80^{\circ}$,~$100^{\circ}$-$180^{\circ}$\\
           & IS   & $0.03$-$0.14$           & $0.04$   &$<45^{\circ}$($> 45^{\circ}$)        &  $\alpha$=$60^{\circ}$-$120^{\circ}$\\
 \hline
VIE (IXE)  & NS         & $0.001$-$0.07$          & $0.016$  &-        &  $\alpha$=$40^{\circ}$-$140^{\circ}$ \\
\hline 
VIF (IXD)  & NS         & $0.002$-$0.30$          & $0.003$ &-        &  -   \\
           & IS         & $0.05$-$0.35$             & $0.044$  &$>45^{\circ}$($< 45^{\circ}$)        &  $\alpha$=$0^{\circ}$-$80^{\circ}$,~$100^{\circ}$-$180^{\circ}$ \\
\hline 
VIID (VIIIF) & NS         & $0$-$0.30$            & $0.0008$   &$>45^{\circ}$($< 45^{\circ}$)   &  -  \\
             & IS        & $0.02$-$0.30$          & $0.045$        &$<45^{\circ}$($> 45^{\circ}$)   &  -   \\
\hline 
VIIE (VIIIE) & NS         & $0.008$-$0.03$       & $0.02$        &-   &  $\alpha$=$50^{\circ}$-$130^{\circ}$ \\
\hline
VIIF (VIIID)& NS         & $0.008$-$0.028$        & $0.007$       &-   &  $\alpha$=$30^{\circ}$-$150^{\circ}$ \\
\hline
VIIIA (XA)& IS     & $0.01$-$0.05$          & $0.0008$     &-    &  $\beta$=$70^{\circ}$-$110^{\circ}$  \\
\hline 
VIIIB (XC)  & NS         & $0.003$-$0.25$         & $0.005$     &-     &  $\alpha$=$0^{\circ}$-$20^{\circ}$,~$160^{\circ}$-$180^{\circ}$ \\
            & IS         & $0.01$-$0.25$          & $0.003$     &-     &  -                              \\
\hline 
VIIIC (XB) & NS         & $0.003$-$0.25$          & $0.005$     &-     &  $\alpha$=$0^{\circ}$-$20^{\circ}$,~$160^{\circ}$-$180^{\circ}$ \\
           & IS         & $0.003$-$0.30$          & $0.003$     &-     &  - \\
\hline
IXA        & IS   & -                     & -         &-     &  - \\
 \hline
IXB (IXC)   & NS        & $0.003$-$0.30$          & $0.0052$    & $36^{\circ}$-$48^{\circ}$($42^{\circ}$-$56^{\circ}$)     &  $\alpha$=$0^{\circ}$-$30^{\circ}$,~$150^{\circ}$-$180^{\circ}$ \\
            & IS  & $0.01$-$0.35$           & $0.0065$    & $36^{\circ}$-$46^{\circ}$($44^{\circ}$-$56^{\circ}$)     &  -  \\
\hline
XD (XF)    & NS         & $0.001$-$0.045$          & $0.0015$   &-      &  $\alpha$=$20^{\circ}$-$160^{\circ}$ \\
           & IS   & $0.03$-$0.13$          & $0.022$    &$<45^{\circ}$($> 45^{\circ}$)      &  $\alpha$=$35^{\circ}$-$145^{\circ}$ \\
\hline  
XE         & NS         & $0.001$-$0.30$              & $0.002$    &-      &  -  \\
           & IS   & -                     & -        &-      &  -   \\
\hline  
\end{tabular}
}
\caption{The predictions for the phenomenologically viable textures. $m_o$ is the lowest mass scale \textit{i.e.} lowest value of $m_1$ for NS and $m_3$ for IS.}
\end{table}
Main results of the numerical analysis are: 
\begin{itemize}
\item Six textures, viz.,\\
$IIA$, $IIIA$, $IVA$, $VA$, $VIA$ and $VIIA$ are excluded by the present experimental data. 
\item Textures $IE$, $IIE$, $IIID$, $IIIE$, $IVE$, $IVF$, $VIE$, $VIIE$, $VIIF$, $VIIID$,$VIIIE$ and $IXE$ lead to a normal spectrum only.
\item Textures $IA$, $IB$, $IC$, $IIC$, $IIIB$, $IIIC$, $IVB$, $VIIIA$, $IXA$ and $XA$ lead to an inverted spectrum only.
\item The allowed points for the following textures are very few for\\
$IB$, $IF$, $IID$, $IIF$, $IIIF$, $IVD$, $VE$, $VID$, $VIIIA$, $IXA$, $IXF$, $XD$, $XE$, $XF$ for an inverted spectrum \\
 and $IXB$, $IXC$ for both inverted and normal spectra. \\
We have generated $10^8$ random points for these textures.
\item All the viable textures except $IA$, $IB$, $IC$, $VIIE$, $VIIIA$, $VIIIE$ and $XA$ allow quasi-degenerate spectrum.
\item Many of the classes predict a constrained range for $M_{ee}$.
\item It is found that the smallest neutrino mass cannot be zero for any of the allowed textures.% except for classes $VIID$(NS) and $VIIIF$(NS). However since we are considering type-I seesaw mechanism and $M_D$ is diagonal, this will lead to $M_R$ with a vanishing eigenvalue and seesaw is not applicable in that limit. So the limiting case of a vanishing eigenvalue for these classes cannot be achieved in the framework of type-I seesaw mechanism in the diagonal $M_D$ basis.
\item For textures:\\
$IIB$, $IIC$, $IIIB$, $IIIC$, $IVB$, $IVC$, $VB$, $VC$, $VD$, $VE$, $VF$, $VIB$, $VIC$, $VIIB$, $VIIC$, $VIIIB$, $VIIIC$, $IXB$, $IXC$, $XB$, $XC$  and  $VIF$, a non-vanishing reactor mixing angle is an inherent property since for $\theta_{13} = 0$ the solar mass square difference ($\Delta m_{21}^{2}$) vanishes \textit{i.e.} $m_1 = m_2$, which is contrary to the experimental observations.
\end{itemize}
We have summarised the numerical results for all the classes compatible with the present experimental data in Table 3. Some of the interesting results are plotted in Figs. 1-5. Fig. 1(a) shows the correlation plot between the two Majorana-type phases ($\alpha$ and $\beta$) for class $IA$(IS). For class $IIB$(NS), $\delta$ approaches $0^o$ or $180^o$ with the increase in $M_{ee}$ [Fig. 1(b)]. Also, for the normal spectrum, $J_{CP}$ cannot be equal to zero [Fig. 1(c)] implying that CP is necessarily violated. There is an interesting correlation between $\theta_{12}$ and $\theta_{23}$ for class $IIIF$(NS) [Fig. 1(d)] where $\theta_{23}$ remains above maximal when $\theta_{12}$ is below $34^o$. For class $IVD$, $\theta_{23}$ and $\theta_{13}$ are correlated [Figs. 1(e)(NS), 1(f)(IS)]. In Fig. 2, we have given correlation plots for classes $VB$ and $VC$ which are among the most predictive textures in this analysis. The Majorana phases are restricted to a very small range for class $VB$(IS) [Fig. 2(a)]. The Dirac-type phase is restricted near $90^o$ or $270^o$ [Fig. 2(b)] and $J_{CP}$ is always non-zero [Fig. 2(c)] implying that, for inverted spectrum, this class is necessarily CP-violating. For class $VC$(NS), $\theta_{23}$ is always above $45^o$ [Fig. 2(d)] and the Dirac-type phase is fixed near $90^o$ and $270^o$ for $M_{ee}>0.05$ eV [Fig. 2(e)]. For the inverted spectrum for this class, $\delta$ again approaches $90^o$ and $270^o$ with increasing $M_{ee}$ [Fig. 2(f)]. Fig. 3(a) depicts the correlation plot between $\delta$ and $M_{ee}$ for class $VE$(IS). For class $VIB$, the Dirac-type phase $\delta$ approaches $0^o$ and $180^o$ with increasing $M_{ee}$ for both normal and inverted spectra [Figs. 3(b) and 3(c)]. Fig. 3(d) shows the correlation plot between $\theta_{23}$ and $\theta_{13}$  for class $VID$(IS) for which $\theta_{23}$ remains below maximal for the allowed ranges of other neutrino oscillation parameters. The correlation plots between $\delta$ and $M_{ee}$ for classes $VIE$ and $VIF$ for normal and inverted spectra are shown in Figs. 3(e)(NS) and 3(f)(IS). For class $VIIF$(NS), $M_{ee}$ is restricted to a small range $0.008$-$0.028$ [Fig. 4(a)] and the correlation plot depicting normal spectrum for this class is shown in Fig. 4(b). For class $VIIIA$(IS), the atmospheric mixing angle $\theta_{23}$ remains near maximal as shown in Fig. 4(c). Fig. 4(d) shows the correlation plot of $\delta$ and $M_{ee}$ for class $VIIIB$(NS) where for larger values of $M_{ee}$, the Dirac phase $\delta$ is restricted to very small ranges. Again, for class $VIIIC$(NS), $\delta$ is restricted to very small ranges for larger values of $M_{ee}$ [Fig. 4(e)]. Fig. 4(f) shows the correlation plot between $\delta$ and $\alpha$ where $\alpha$ has a small allowed range near $0^o$ and $180^o$ [Fig. 4(f)]. The $2$-$3$ interchange symmetry between classes $VIIIF$ and $VIID$ for NS is shown in Figs. 5(a) and 5(b).
In addition to the textures for which non-zero $\theta_{13}$ is an inherent property, there are other textures namely $IA$, $VIIIA$, $IXA$ and $XA$ which forbid a vanishing $\theta_{13}$. This is because for $\theta_{13} = 0$, these textures predict 
\begin{equation}
\left|\frac{m_1}{m_2}\right| = \left|-\cot^2 \theta_{12}\right| > 1 \ \ \ \textrm{for the allowed range of $\theta_{12}$}
\end{equation}
which is contrary to the experimental observations.

\begin{figure}
{\epsfig{file=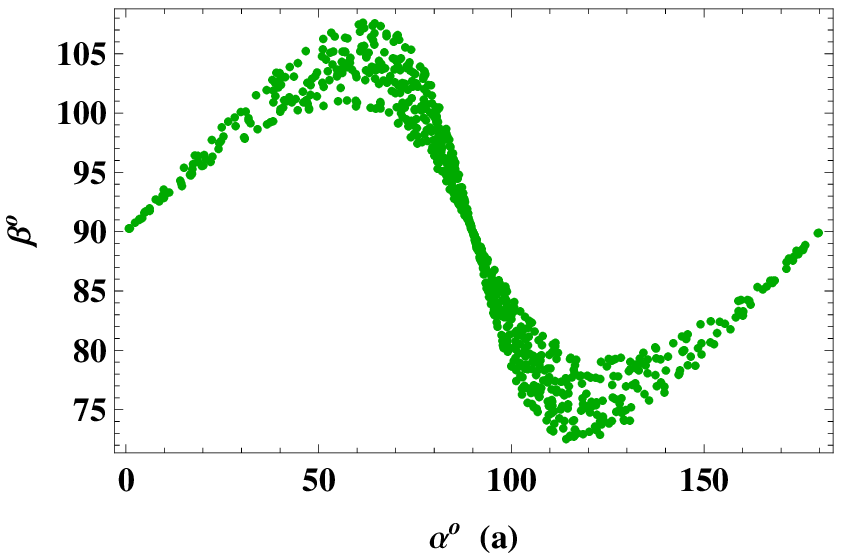, width=5.0cm, height=4.0cm} 
\epsfig{file=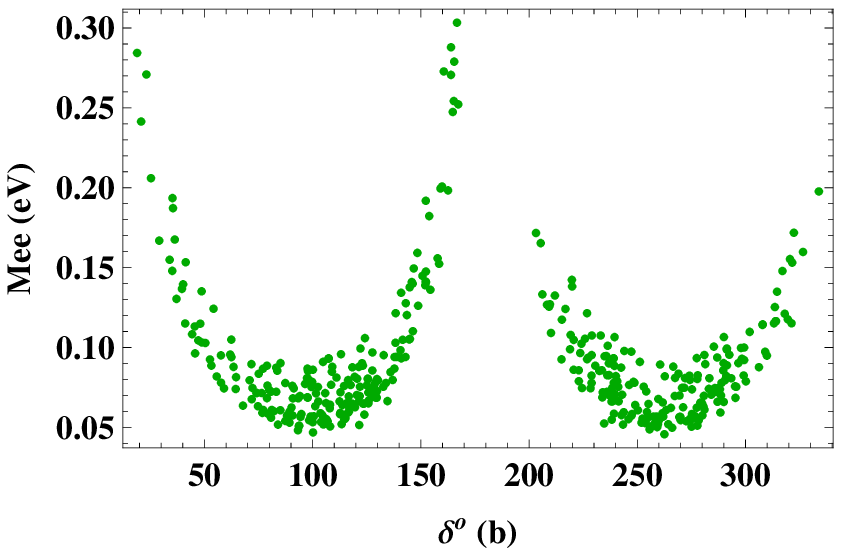, width=5.0cm, height=4.0cm} 
\epsfig{file=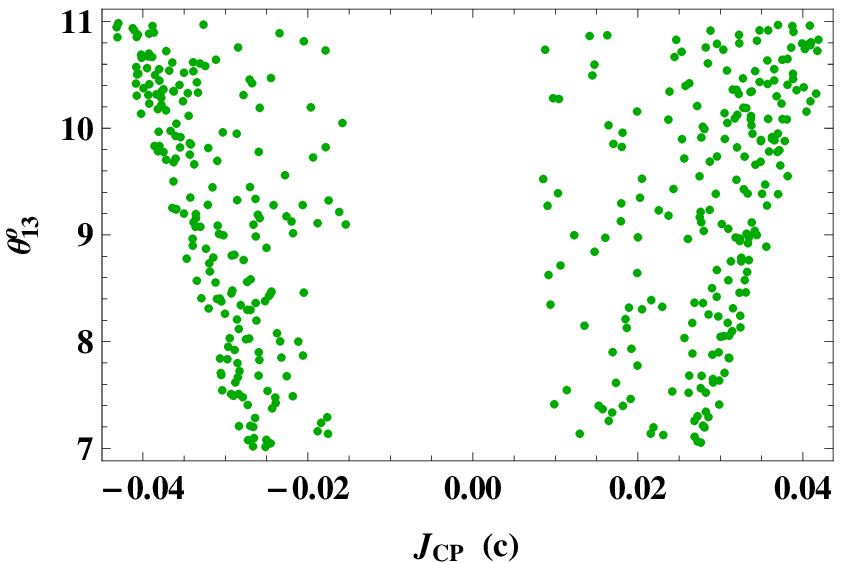, width=5.0cm, height=4.0cm} \\ 
\epsfig{file=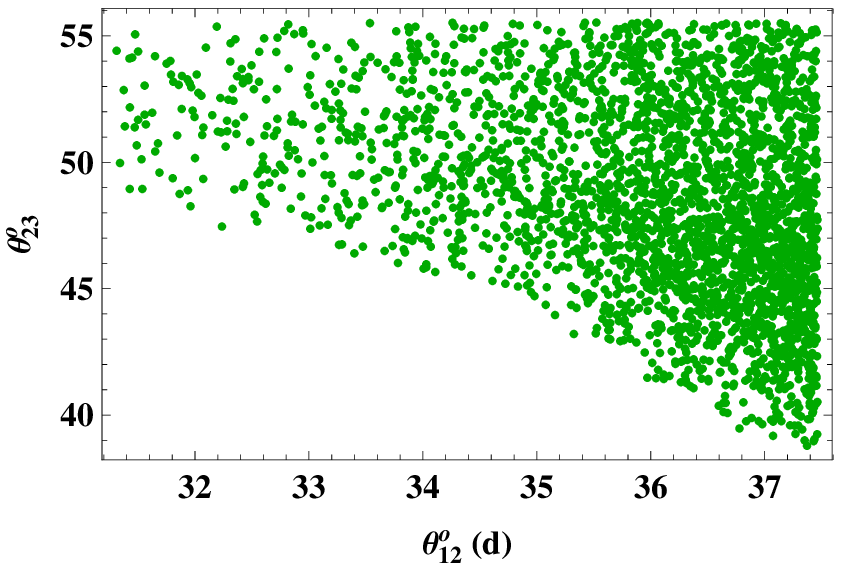, width=5.0cm, height=4.0cm} 
\epsfig{file=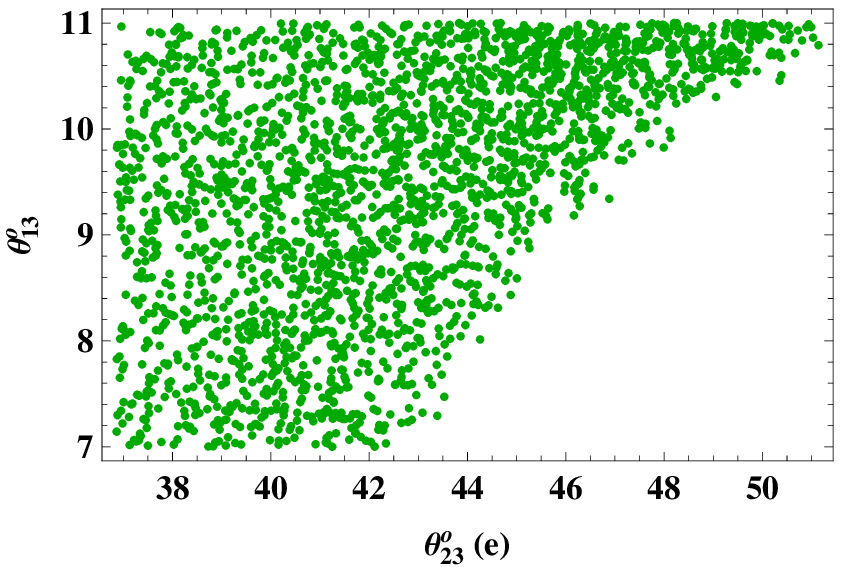, width=5.0cm, height=4.0cm}  
\epsfig{file=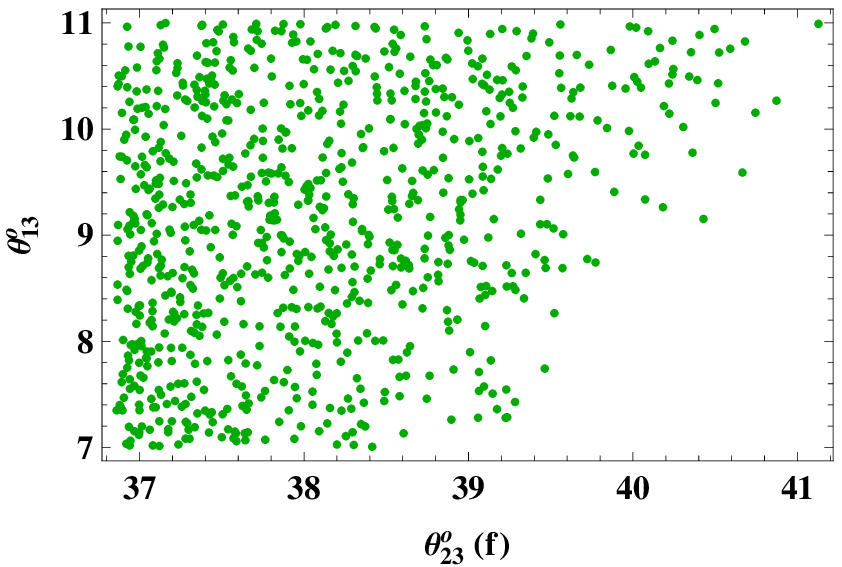, width=5.0cm, height=4.0cm}}
\caption{Correlation plots for classes $IA$(IS)(a), $IIB$(NS) (b, c), $IIIF$(NS)(d), $IVD$(NS)(e) and $IVD$(IS)(f).}
\end{figure}
 
\begin{figure}
{\epsfig{file=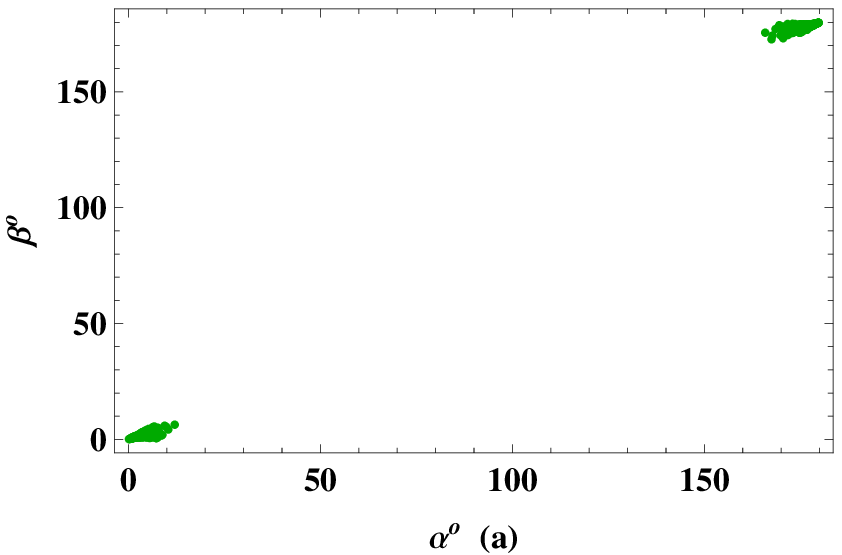, width=5.0cm, height=4.0cm}  
\epsfig{file=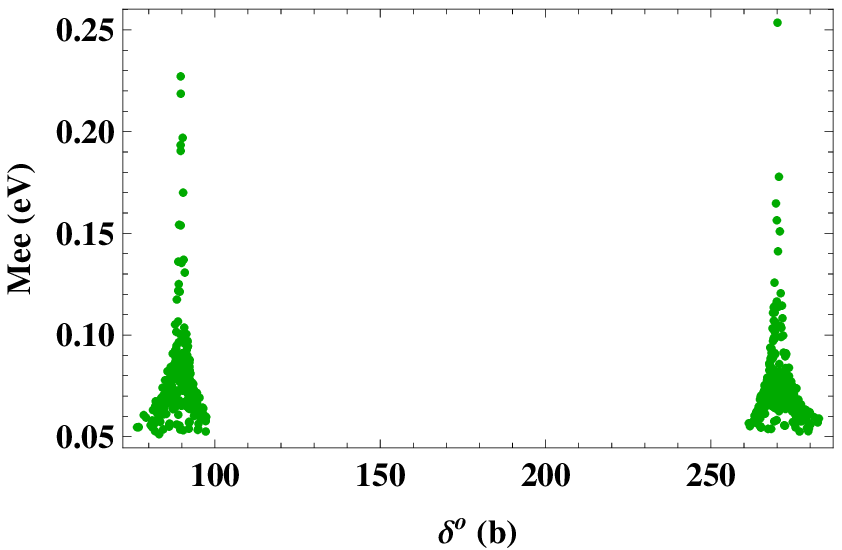, width=5.0cm, height=4.0cm} 
\epsfig{file=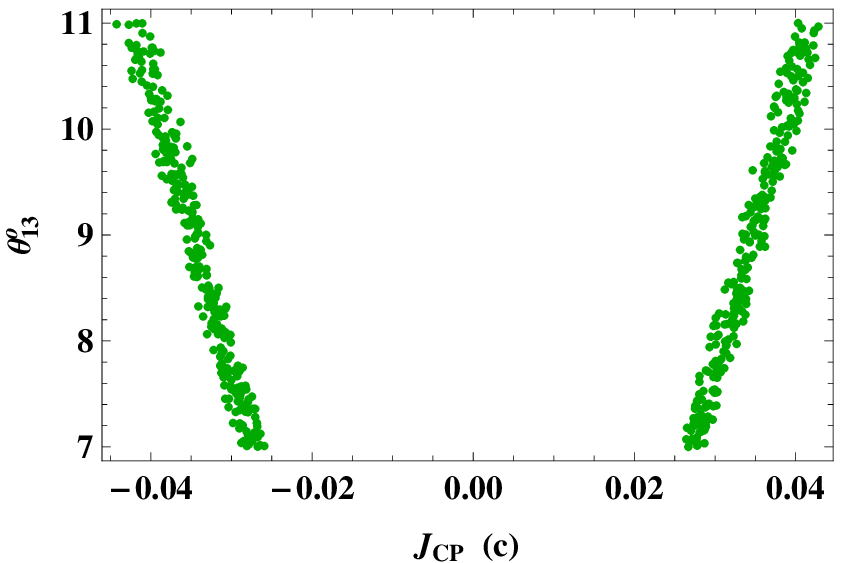, width=5.0cm, height=4.0cm}\\   
\epsfig{file=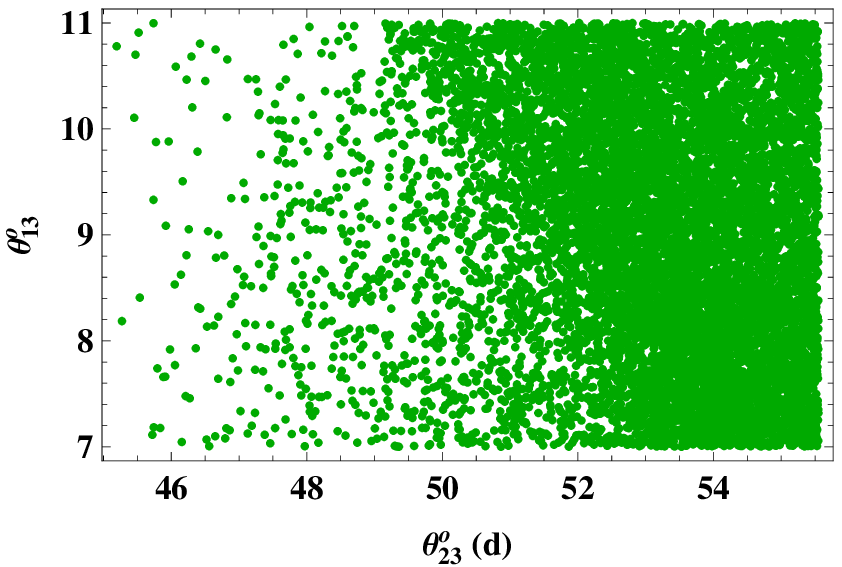, width=5.0cm, height=4.0cm}  
\epsfig{file=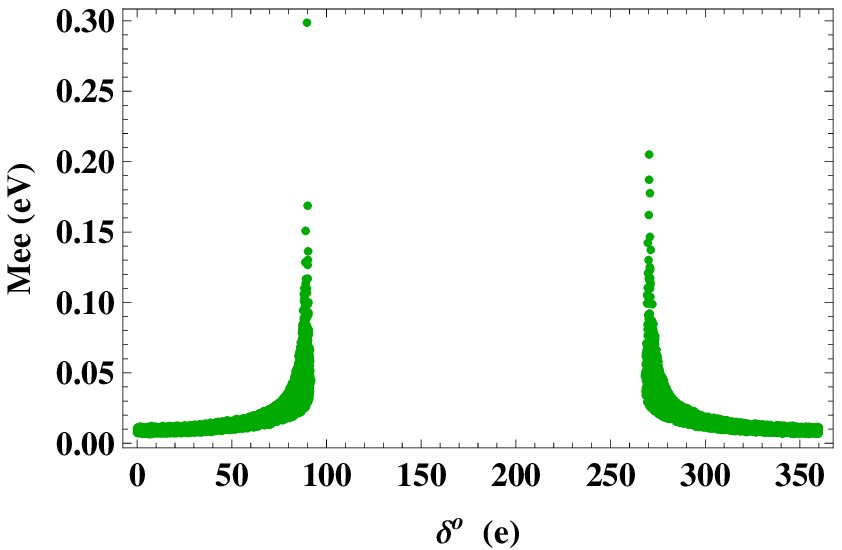, width=5.0cm, height=4.0cm}  
\epsfig{file=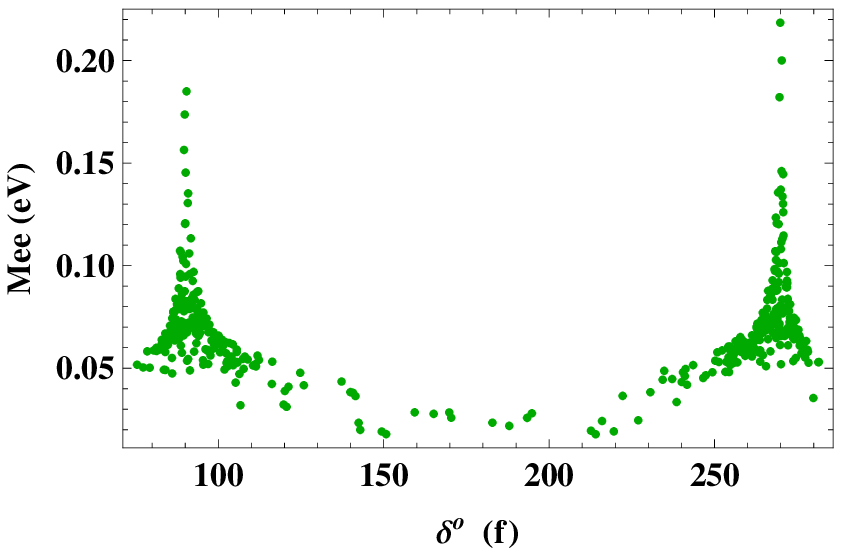, width=5.0cm, height=4.0cm}}
\caption{Correlation plots for classes $VB$(IS)(a, b, c), $VC$(NS) (d, e) and $VC$(IS)(f).}
\end{figure}

\begin{figure} 
{\epsfig{file=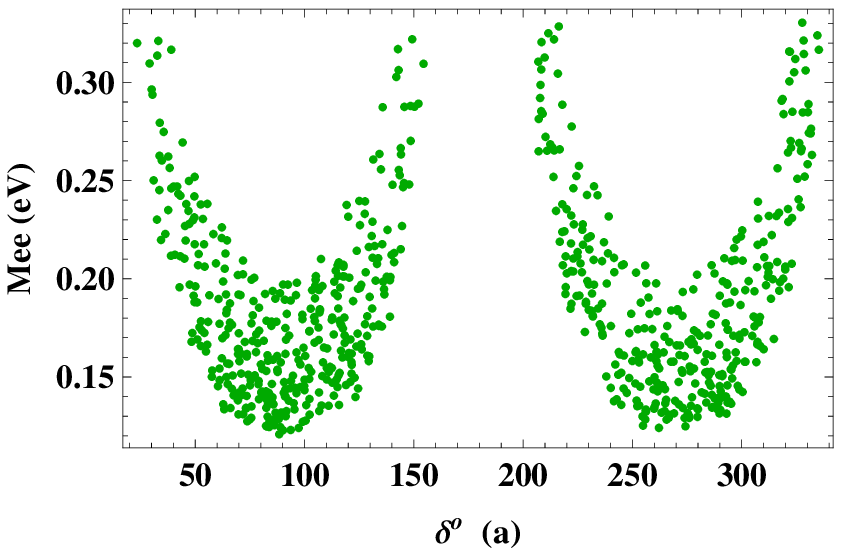, width=5.0cm, height=4.0cm}  
\epsfig{file=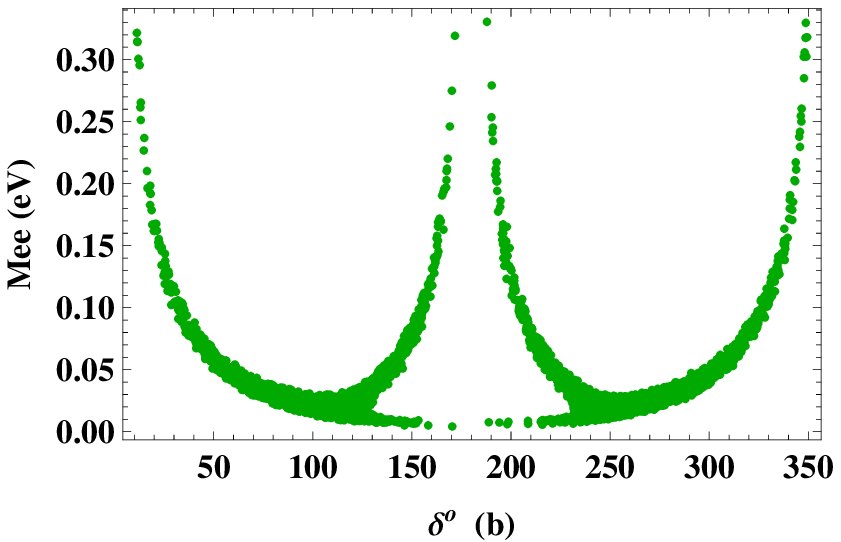, width=5.0cm, height=4.0cm}  
\epsfig{file=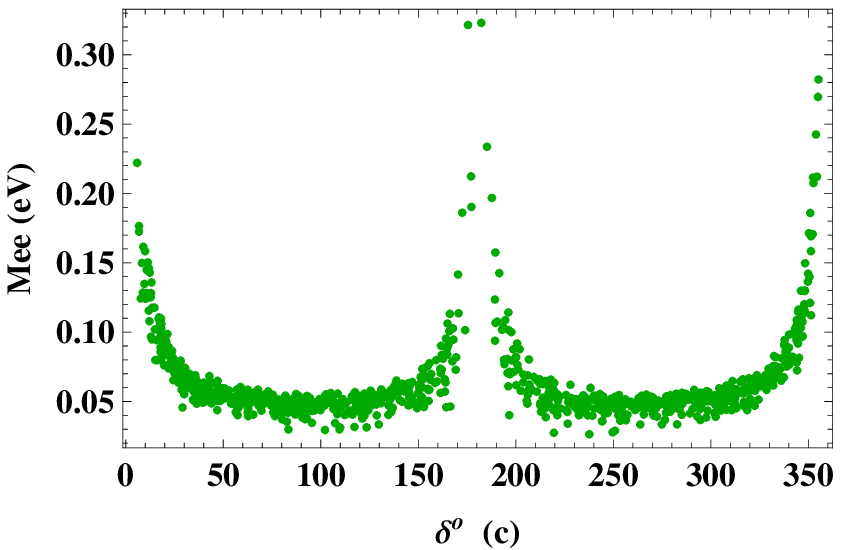, width=5.0cm, height=4.0cm}\\  
\epsfig{file=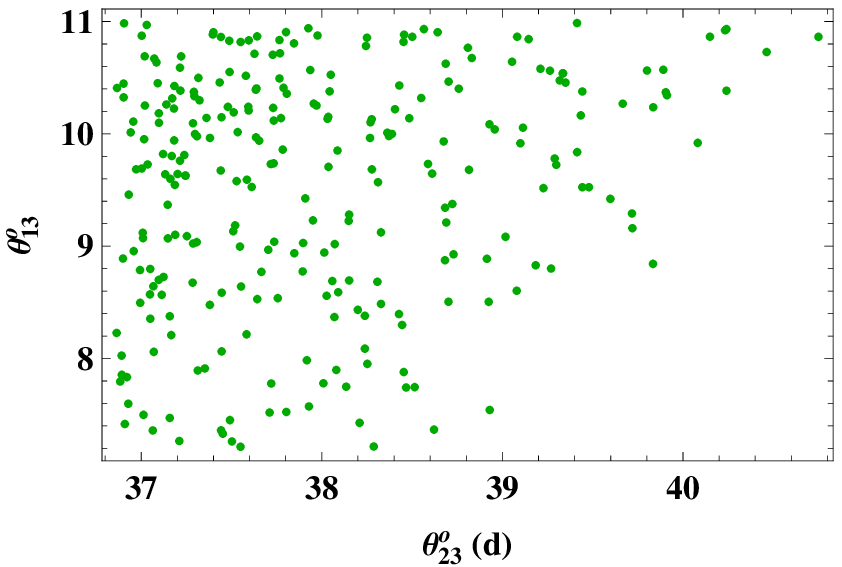, width=5.0cm, height=4.0cm}
\epsfig{file=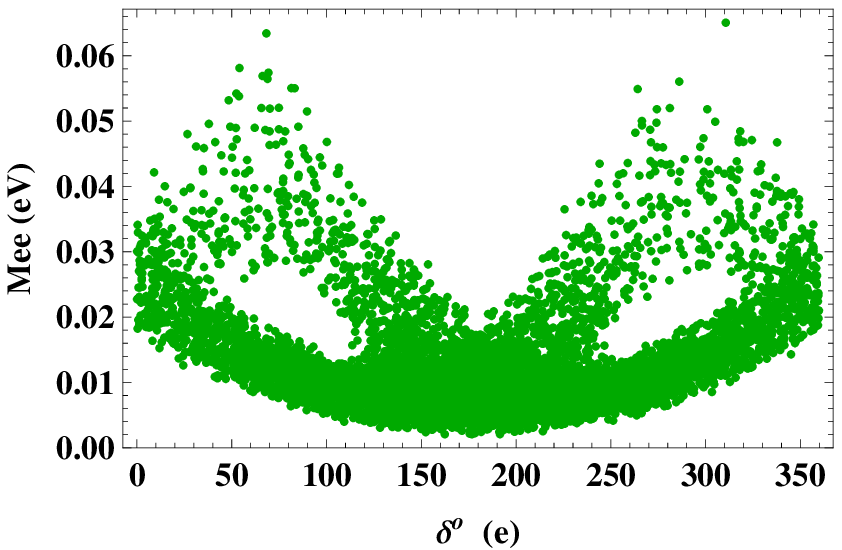, width=5.0cm, height=4.0cm}   
\epsfig{file=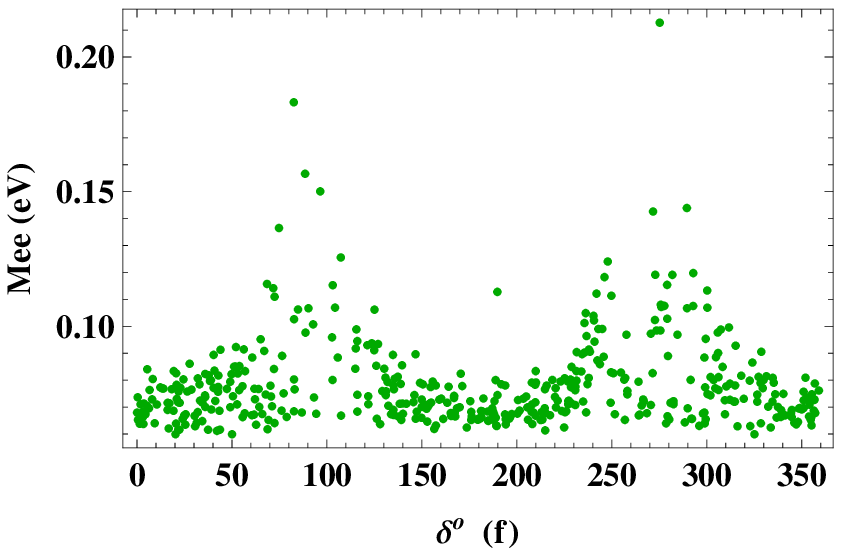, width=5.0cm, height=4.0cm}}
\caption{Correlation plots for classes $VE$(IS)(a), $VIB$(NS)(b), $VIB$(IS)(c), $VID$(IS)(d), $VIE$(NS)(e) and $VIF$(IS)(f).}
\end{figure} 

\begin{figure}
{\epsfig{file=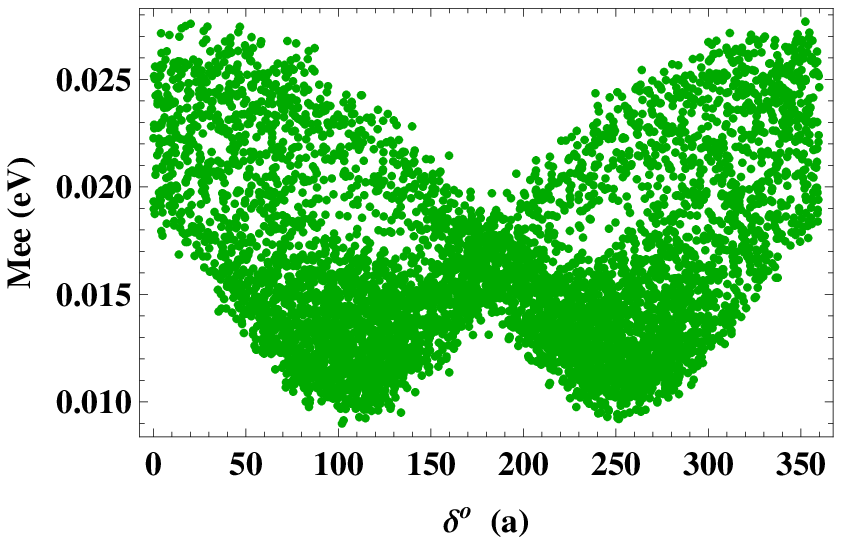, width=5.0cm, height=4.0cm}  
\epsfig{file=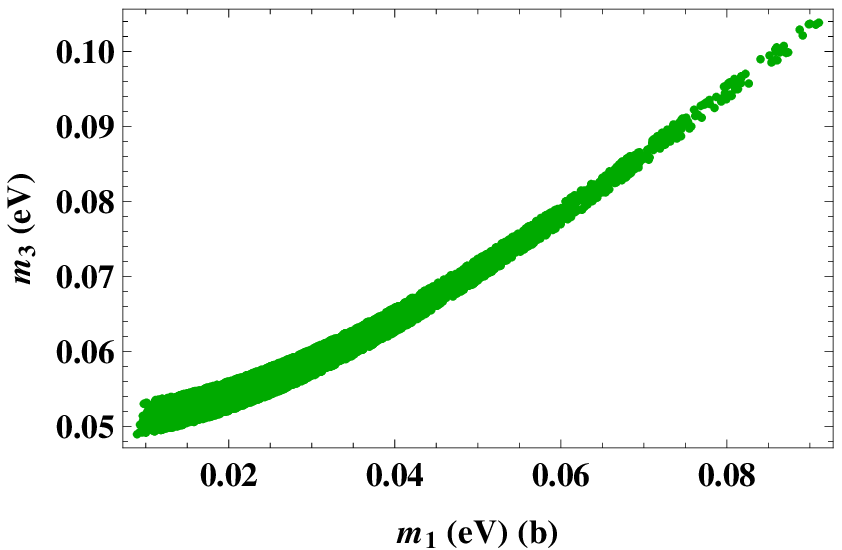, width=5.0cm, height=4.0cm}
\epsfig{file=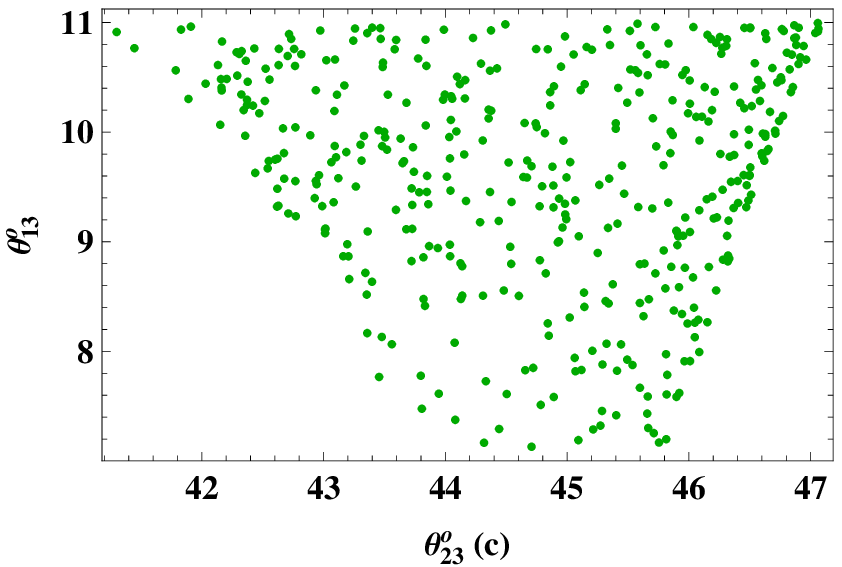, width=5.0cm, height=4.0cm}\\ 
\epsfig{file=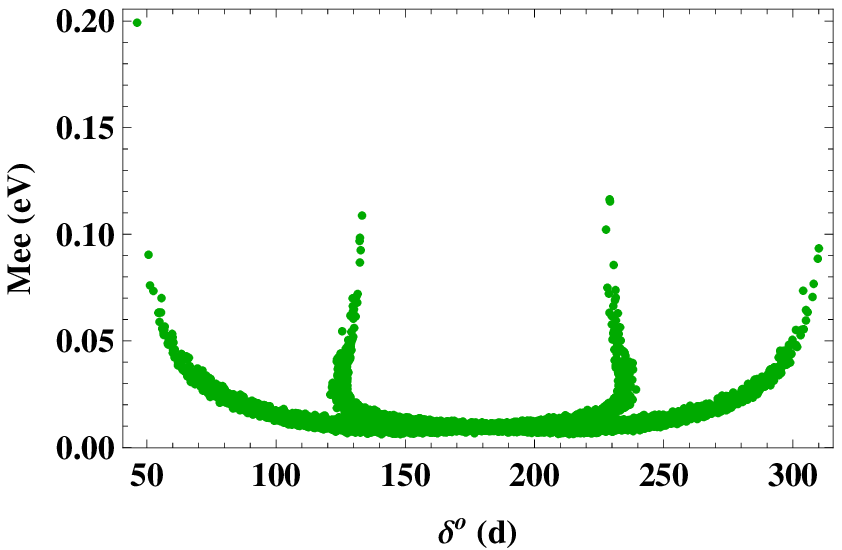, width=5.0cm, height=4.0cm}  
\epsfig{file=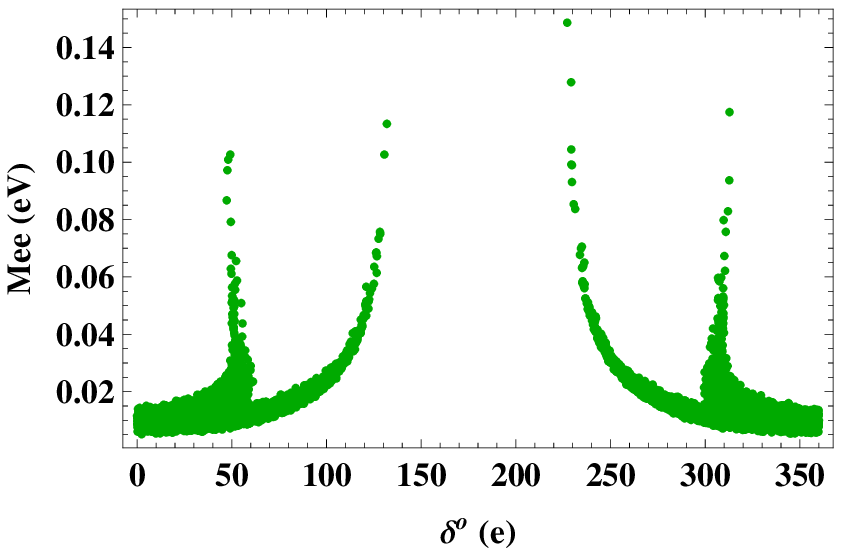, width=5.0cm, height=4.0cm}  
\epsfig{file=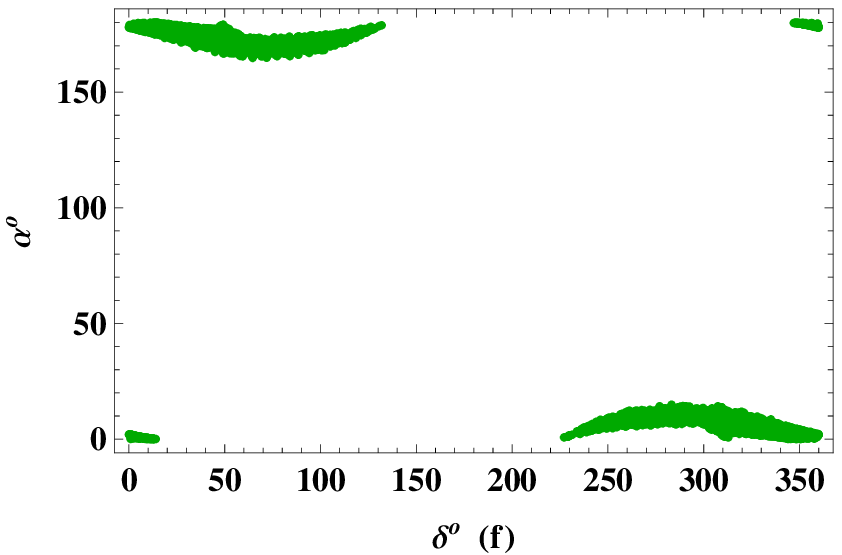, width=5.0cm, height=4.0cm}}
\caption{Correlation plots for classes $VIIF$(NS)(a, b), $VIIIA$(IS)(c), $VIIIB$(NS)(d) and $VIIIC$(NS)(e, f).}
\end{figure}

\begin{figure}
\begin{center}
{\epsfig{file=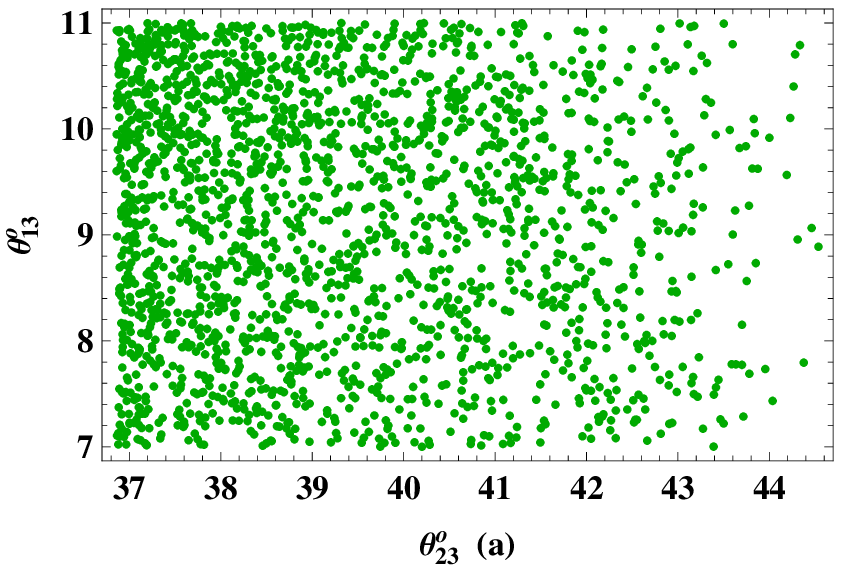, width=5.0cm, height=4.0cm}
\epsfig{file=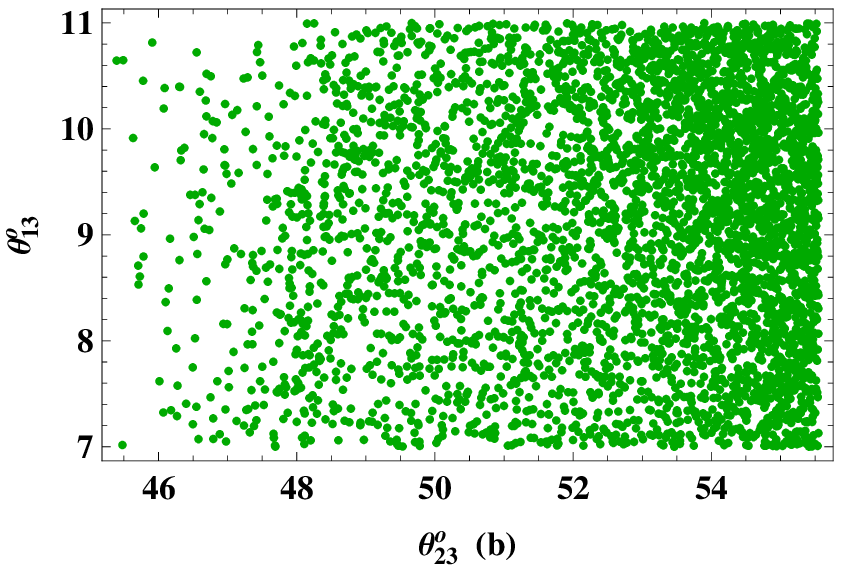, width=5.0cm, height=4.0cm}  
}
\caption{Correlation plots for classes $VIIIF$(NS)(a) and $VIID$(NS)(b) depicting the $2$-$3$ interchange symmetry.}
\end{center}
\end{figure}

\section{Symmetry Realization}
We present a representative model for obtaining hybrid textures in the right-handed neutrino mass matrix $M_R$. Here we need one equality and one vanishing entry in $M_R$ in a basis where both $M_D$ and $M_l$ are diagonal with additional equality in $M_D$. The additional equality in $M_D$ is needed so that the equal elements of $M_R$ may propagate as equal cofactors in $M_\nu$. General guidelines for obtaining zero entries at any place in fermion mass matrices using Abelian discrete symmetries have been propounded in \cite{abeliansym}. In particular, all the viable cases of two zero textures in the neutrino mass matrix in flavor basis have been realized in \cite{xingzt2011, grimustz} considering the type-II seesaw contribution and in \cite{ourtz} considering both type-(I+II) seesaw contributions and using a minimal cyclic symmetry group. A few cases of hybrid textures in the effective neutrino mass matrix have been realised in \cite{zhouhybrid, q8}. Here we present the symmetry realization of class $IA$ as an illustration of how the hybrid textures of $M_\nu^{-1}$ may be realised. For this we follow the approach of Ref. \cite{grimusantimutau}.\\
In addition to the Standard Model (SM) left-handed $SU(2)$ lepton doublets $D_{\ell L}$ $(\ell = e, \mu, \tau)$ and the right-handed charged-lepton $SU(2)$ singlets $\ell_R$, we introduce three right-handed neutrinos $\nu_{\ell R}$. In the scalar sector we need three Higgs doublets $\phi_\jmath$ ($\jmath=1,2,3$), we will also need three scalar singlets. We assume the validity of the lepton number symmetries $U(1)_{L_\ell}$  for all dimension 4 terms in the Lagrangian. This leads to diagonal charged lepton $(M_l)$ and Dirac neutrino $(M_D)$ mass matrices. However, the dimension 3 terms are allowed to break the $L_\ell$ softly.\\
We consider the following symmetries and transformations of various fields under these symmetries:
\begin{align}
 \mathbbm{Z}_2:\hspace{1.5cm}  \phi_1, e_R, \nu_{\ell R}  &\longrightarrow  -\phi_1, -e_R, -\nu_{\ell R} \\ \nonumber
\\ 
\mathbbm{Z}_4: \hspace{2.4cm} D_{eL}&\longrightarrow -i D_{eL},& D_{\mu L}&\longrightarrow i D_{\tau L},& D_{\tau L}&\longrightarrow i D_{\mu L},& \nonumber \\e_{R}&\longrightarrow -i e_{R},& \mu_{R}&\longrightarrow i \tau_{R},& \tau_{R}&\longrightarrow i \mu_{R},& \\ \nu_{e R}&\longrightarrow -i \nu_{e R},& \nu_{\mu R}& \longrightarrow  i \nu_{\tau R},& \nu_{\tau R}&\longrightarrow i \nu_{\mu R},& \nonumber  \\  \phi_3 & \longrightarrow -\phi_3 \ . & \nonumber
\end{align}  
These symmetries lead to the following Yukawa Lagrangian:
\begin{align}
-\mathcal{L}_Y = \ & Y_1 (\overline{D}_{e L} e_R) \phi_1 + Y_2(\overline{D}_{\mu L} \mu_R + \overline{D}_{\tau L} \tau_R) \phi_2 + Y_3(\overline{D}_{\mu L} \mu_R - \overline{D}_{\tau L} \tau_R) \phi_3 + Y_4 (\overline{D}_{e L} \nu_{e R}) \tilde{\phi_1} \nonumber \\ \ & + Y_5 (\overline{D}_{\mu L} \nu_{\mu R} + \overline{D}_{\tau L} \nu_{\tau R}) \tilde{\phi_1} + \ \textrm{H. c.}
\end{align}
where $\tilde{\phi_1} = i \tau_2 \phi_1^*$.
When the Higgs fields ($\phi_\jmath$) acquire non-zero vacuum expectation values (VEVs), we have the charged lepton mass matrix $M_l$ and the Dirac neutrino mass matrix $M_D$ of the following form:
 \begin{align}
 M_l &= \textrm{diag}(e^{i \phi_e} m_e, e^{i \phi_\mu} m_\mu, e^{i \phi_\tau} m_\tau)\\
 M_D &= \textrm{diag}(x,y,y).
 \end{align}
where $m_e = |Y_1 \langle \phi_1 \rangle_o|$ , $m_\mu = |Y_2 \langle \phi_2 \rangle_o +  Y_3 \langle \phi_3 \rangle_o|$, $m_\tau = |Y_2 \langle \phi_2 \rangle_o - Y_3 \langle \phi_3 \rangle_o|$, $x = Y_4 \langle \phi_1^* \rangle_o$ and $y = Y_5 \langle \phi_1^* \rangle_o$.\\
The Majorana mass terms for the right-handed neutrinos, invariant under above symmetries are
\begin{align}
\mathcal{L}_M = \ &\frac{M_1}{2} \nu_{e R}^T C^{-1}( \nu_{\mu R} + \nu_{\tau R}) + \frac{M_2}{2}(\nu_{\mu R}^T C^{-1} \nu_{\mu R} - \nu_{\tau R}^T C^{-1} \nu_{\tau R}) + \ \ \textrm{H. c.}
\end{align}
These mass terms lead to a right-handed Majorana neutrino mass matrix having the following form
 \begin{equation}
 M_R = \left(
\begin{array}{ccc}
0 & a & a \\ a & b & 0 \\ a& 0 & -b
\end{array}
\right).
\end{equation}

All the non-zero entries are generated through dimension 3 terms in the Lagrangian for $M_R$. Now we add three complex scalar singlets $\chi_{\mu \tau}$, $\chi_{\mu\mu}$ and $\chi_{\tau\tau}$. The lepton number $U(1)_{L_\ell}$ assignments of these scalar singlets are given in Table 4. Under the action of $\mathbbm{Z}_4$, $\chi_{\mu \mu} \leftrightarrow -\chi_{\tau \tau}$ and $\chi_{\mu \tau}$ remains invariant. These scalar fields generate Yukawa couplings $Y_6[(\nu_{\mu R}^T C^{-1} \nu_{\mu R})\chi_{\mu\mu} + (\nu_{\tau R}^T C^{-1} \nu_{\tau R})\chi_{\tau\tau}]+$ H.c. and $Y_7(\nu_{\mu R}^T C^{-1} \nu_{\tau R})\chi_{\mu\tau}+$ H.c.. When the scalar singlet fields acquire non-zero VEVs one gets the desired form of $M_R$ for class $IA$:
 \begin{equation}
 M_R = \left(
\begin{array}{ccc}
0 & a & a \\ a & c+b & d \\ a& d & c-b
\end{array}
\right).
\end{equation}
\begin{table}[h]
\begin{center}
\begin{tabular}{|c|c|c|c|c|c|c|c|c|}
\hline  & $\chi_{\mu\tau}$ & $\chi_{\mu\mu}$ & $\chi_{\tau\tau}$ \\
\hline $L_e$ & 0 & 0 & 0 \\ 
\hline $L_\mu$ & -1 & -2 & 0 \\ 
\hline $L_\tau$ & -1 & 0 & -2 \\ 
\hline 
\end{tabular}
\end{center}
\caption{Lepton number assignments of the scalar singlets.}
\end{table} 

\section{Summary}
We presented a detailed phenomenological analysis of neutrino mass matrices having one equality between cofactors and one vanishing cofactor. Such texture structures arise via type-I seesaw mechanism when the Dirac neutrino mass matrix is diagonal with one equality between the elements and the right-handed Majorana neutrino mass matrix has hybrid textures. Out of the total sixty possible hybrid textures in the right-handed Majorana neutrino mass matrix, only six are disallowed by the present neutrino oscillation data. Many of the allowed textures have constrained ranges for the unknown parameter $M_{ee}$, which will be probed in many forthcoming experiments for neutrinoless double beta decay. Some of the allowed textures have predictions for the Dirac-type CP-violating phase $``\delta"$. Many textures also predict interesting correlations between $M_{ee}$ and $\delta$. In addition, there are predictions for the quadrant of the atmospheric mixing angle and we have compiled this information in Table 3. To show how such texture structures can be derived, we presented a flavor model for class $IA$ using discrete flavor symmetries. At present, most of the hybrid textures of the inverse neutrino mass matrix can accommodate the available experimental data. Since most of the textures studied in this analysis have predictions for one or more presently unknown neutrino parameters, experimental results on these neutrino parameters such as the quadrant of $\theta_{23}$, the value of Dirac-type CP-violating phase and the magnitude of the effective Majorana mass $M_{ee}$ will help in deciding the viability of these textures in explaining neutrino masses and mixings. \\

Note: After the completion of this work a similar analysis with somewhat different results came to our notice \cite{wang}. \\

\textbf{\textit{\Large{Acknowledgements}}}\\
R. R. G. acknowledges the financial support provided by the Council for Scientific and Industrial Research (CSIR), Government of India.

\end{document}